\documentclass[%
 aip,
 amsmath,amssymb,twocolumn,
reprint,%
]{revtex4-1}

\usepackage{graphicx}
\usepackage{dcolumn}
\usepackage{bm}
\usepackage{amsmath}
\usepackage{color}
\usepackage{epstopdf}
\usepackage{float}
\usepackage{bibentry}
\usepackage{dcolumn}

\begin{document}

\preprint{AIP/123-QED}

\title{Characterization of hidden modes in networks of superconducting qubits}

\author{Sarah Sheldon}
\author{Martin Sandberg}
\author{Hanhee Paik}
\author{Baleegh Abdo}
\author{Jerry M. Chow}
\author{Matthias Steffen}
\author{Jay M. Gambetta}
\affiliation{IBM T.J. Watson Research Center, Yorktown Heights, NY 10598, USA}


\date{\today}


\newcommand{\ket}[1]{\left|#1\right>}
\newcommand{\bra}[1]{\left<#1\right|}
\newcommand{\bracket}[2]{\left<#1|#2\right>}
\newcommand{\Bracket}[3]{\left<#1|#2|#3\right>}
\newcommand{\expvalue}[2]{\left<#1|#2|#1\right>}

\begin{abstract}
We present a method for detecting electromagnetic (EM) modes that couple to a superconducting qubit in a circuit quantum electrodynamics (circuit-QED) architecture.  Based on measurement-induced dephasing, this technique allows the measurement of modes that have a high quality factor (Q) and may be difficult to detect through standard transmission and reflection measurements at the device ports.  In this scheme the qubit itself acts as a sensitive phase meter, revealing modes that couple to it through measurements of its coherence time.  Such modes are indistinguishable from EM modes that do not couple to the qubit using a vector network analyzer. Moreover, this technique provides useful characterization parameters including the quality factor and the coupling strength of the unwanted resonances.  We demonstrate the method for detecting both high-Q coupling resonators in planar devices as well as spurious modes produced by a 3D cavity.
\end{abstract}

\maketitle


Superconducting qubits within a circuit-QED architecture are a prime candidate for quantum computing devices. Qubit coherence times have exceeded\cite{dial} 100~$\mu$s, and gate fidelities have reached  over 0.999 \cite{mckay, rol,kelly} for single- and 0.99 for two-qubit operations \cite{barends,sheldonCR}. Experiments are now moving beyond devices with few qubits and instead involve arrays of connected qubits forming large, complicated networks of qubits and quantum buses or coupling resonators \cite{kelly15, riste, takita}. In these systems, there are many high-Q modes that have the potential to couple to the qubits, either by design in the case of bus resonators, or as a result of spurious modes determined by the geometry of the device and packaging \cite{houck}.   In this letter, we describe a simple method for characterizing these various EM modes by using the qubit as a sensitive phase meter to detect any electromagnetic mode that couples to it.  As we will show here, this technique, which we refer to as ``coherence spectroscopy", identifies only those modes which couple to the qubit and does so with a sensitivity much greater than that of external port $S$-parameter measurements.

\begin{figure*}[!ht]
	\centering
	\begin{align*}
	\includegraphics[width=\columnwidth]{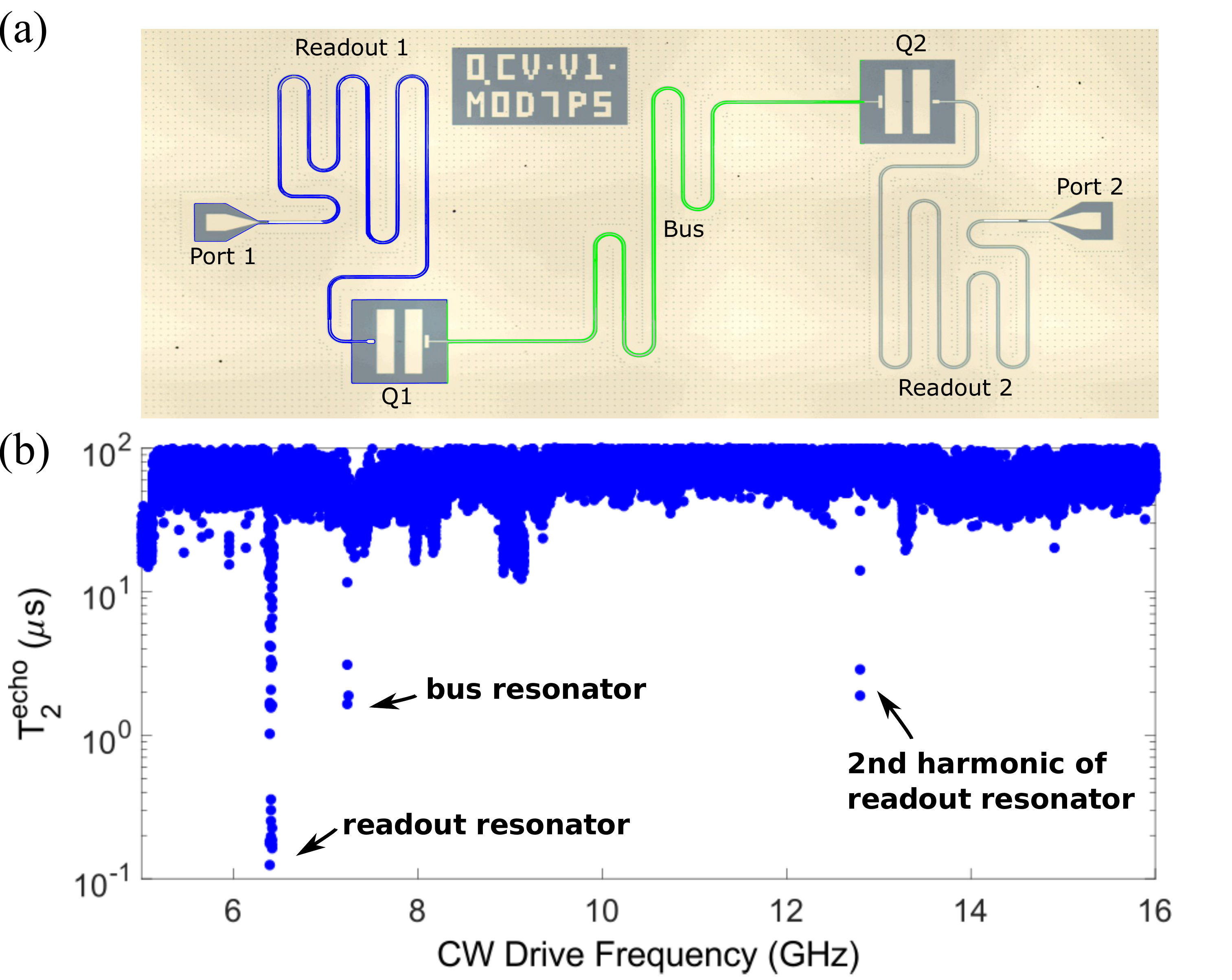} &\includegraphics[width=\columnwidth]{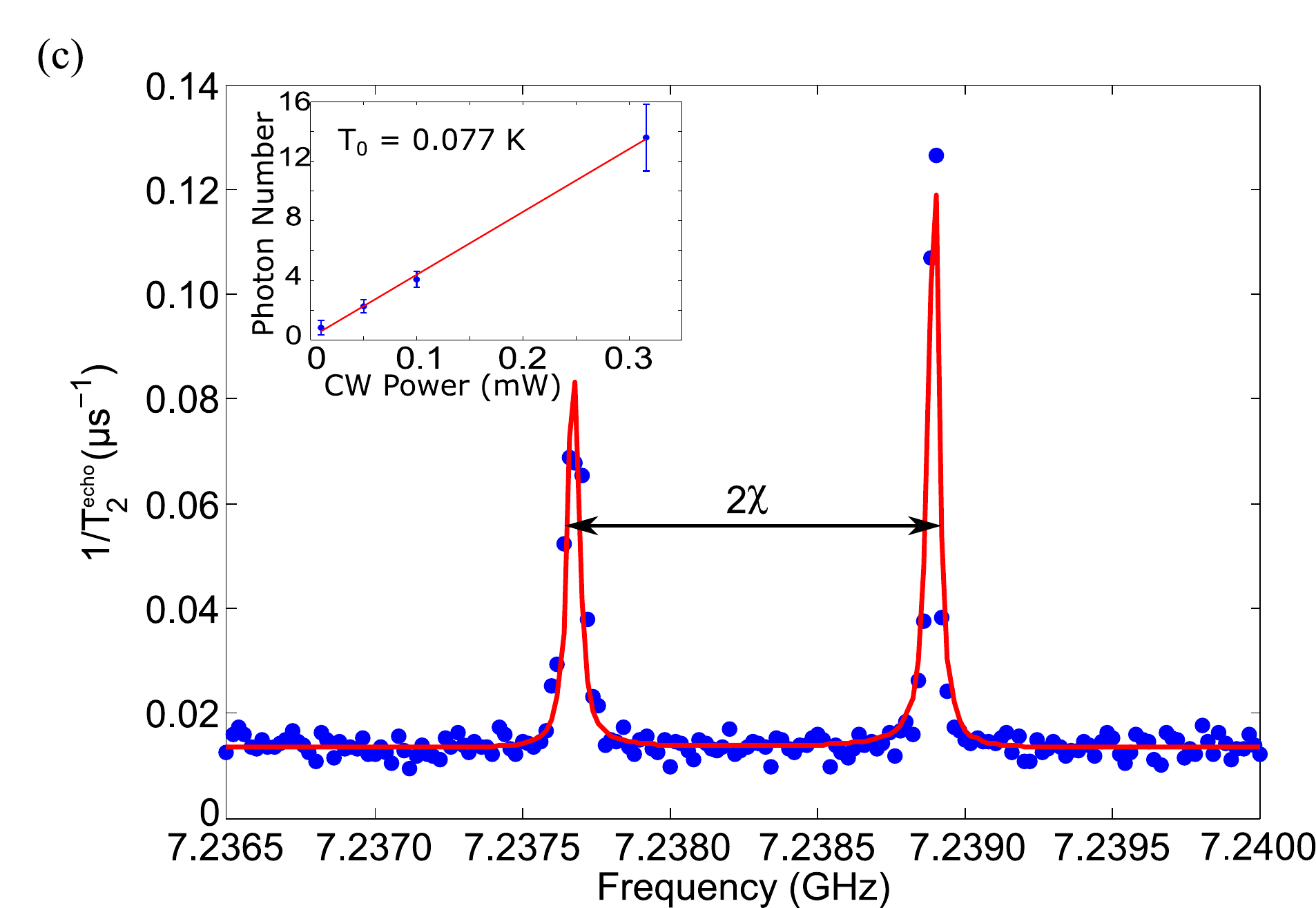}\\		
	\end{align*}
	\caption{(a) Image of the two qubit sample with a single bus resonator (green).  Each qubit has an individual readout resonator (blue and grey). (b) Plot of qubit $T_2^{\mathrm{echo}}$ as a function of the applied CW drive.  The nominal $T_2^{\mathrm{echo}}$ of the qubit is $65\pm11\mu$s but falls to zero the the microwave drive is on resonance with one of the bus cavities. (c) $1/T_2^{\mathrm{echo}}$ versus CW drive frequency on resonance with one bus resonator with the fit to data in red. The inset shows the photon number versus drive power - the linear relationship between the two agrees with theory. $T_0$ is the expected cavity temperature given the photon number at zero CW power extrapolated from the fit.}
	\label{fig:Q1_echoSpect}
\end{figure*}

Typically bus resonators are designed to act as a mechanism for coupling two or more qubits but are difficult to detect through the on-chip measurement ports\cite{majer}. Quantum fluctuations in the number of photons in a resonator coupled to the qubit can create random phase differences between the $\ket{0}$ and $\ket{1}$ state of the qubit, which leads to dephasing.  Measurement-induced-dephasing is a well known phenomenon \cite{schuster,gambetta_measurementInducedDephasing_2006,boissonneault,sears,clerk} that has been characterized in experimental systems using tunable couplings between the qubit and its environment\cite{bertet} and has served as a mechanism for measuring thermal noise\cite{utami, you07, yan2016}. Various efforts have been made to avoid populating resonators in qubit gates that depend on the qubit-resonator coupling \cite{map,rip}, to reduce qubit dephasing through measurement feedback \cite{kockum12, delange, tornberg} or tunable coupling to the readout resonator \cite{zhang}. However, we can also take advantage of this phenomenon to determine the frequency of bus resonators and the strength of bus-qubit coupling.  We accomplish this by accurately measuring the qubit dephasing as a function of the frequency of a field applied to the qubit control port (i.e. Port 1 in Fig \ref{fig:Q1_echoSpect} [a]). The notion of using qubit as a probe for its environment is not new. It has previously been done to study material defects and two level systems (TLS) coupled to the qubit \cite{martinis05,lupascu09,shalibo10}. However our method is different in that it does not require a frequency tunable qubit and it allows us to probe the spectral environment far detuned from the qubit.   

Coherence spectroscopy is not limited to measuring known microwave resonances of the superconducting circuit but can also identify spurious electromagnetic (EM) modes that couple to the qubit. These modes, which can  arise from the sample housing or cavity boundaries in the case of qubits in 3D microwave cavities\cite{paik}, lead to reduced qubit coherence times.  Spurious modes are difficult to characterize as they may couple weakly to the device ports and cannot be distinguished from other microwave modes that do not contribute to qubit dephasing.

\begin{figure}[h]
ert	\includegraphics[width = \columnwidth]{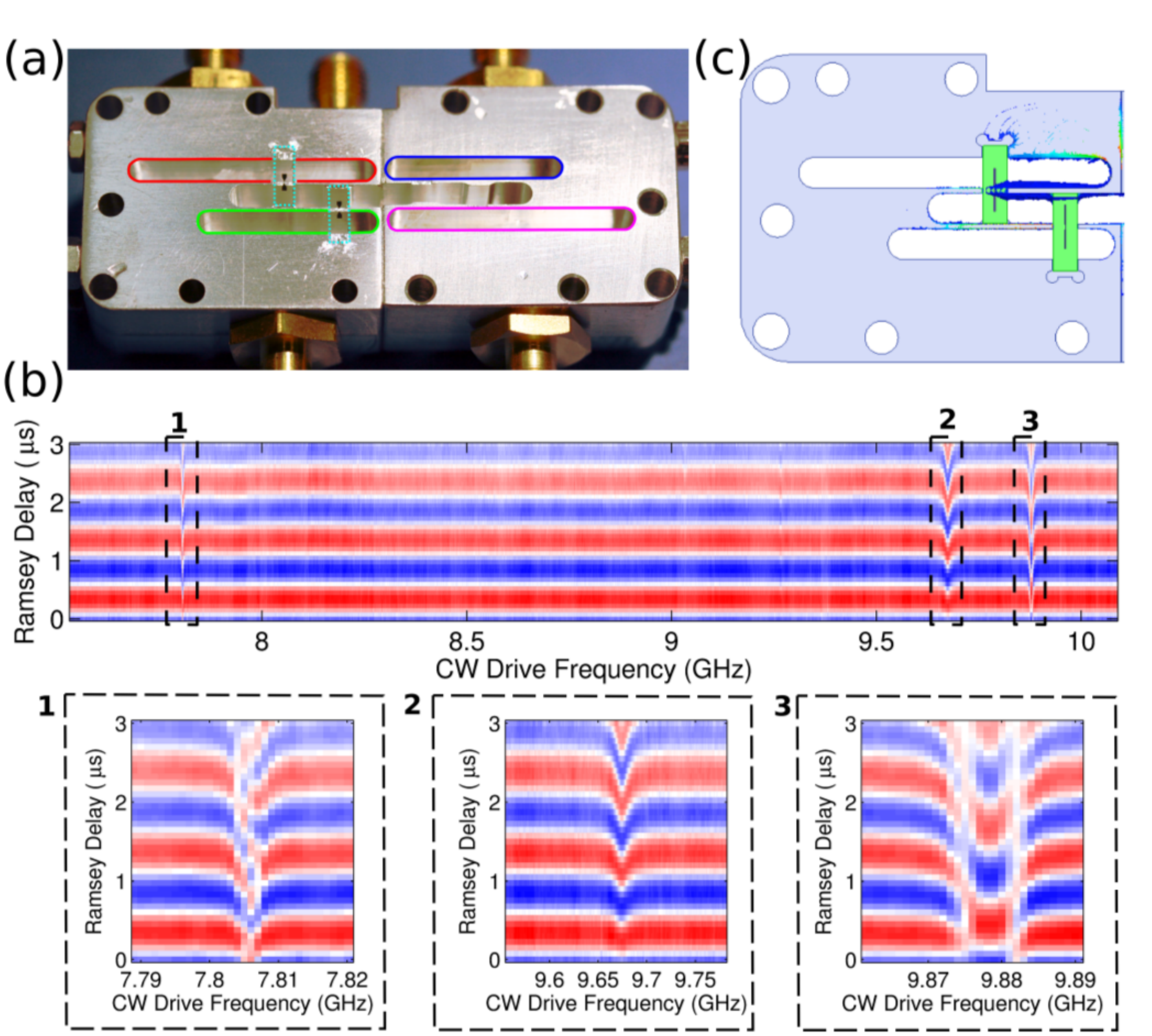}\\
	
	\caption{(a) Image of one half of the enclosure in  the multi-cavity-qubit system. The aluminum enclosure consists of five microwave cavities and four qubit chips. Each qubit is coupled to two cavities: a common bus cavity and an individual readout cavity. The readout cavities ($R1$, $R2$, $R3$ and $R4$) are indicated by blue, red, green and purple outlines respectively in the figure ($R1$ is top right,$R2$ top left, $R3$ bottom left and $R4$ is bottom right). Two qubits ($Q2$ top and $Q3$ bottom) are shown mounted. (b) Color plot of unechoed coherence spectroscopy from 7.5 GHz to 10.2 GHz. Along the y-axis we observe Ramsey oscillations of $Q2$ as we vary the frequency of the CW tone applied to the bus cavity along the x-axis.  We observe three sharp features in the spectrum,$\omega_{\mathbf{1}}/2\pi\approx 7.805$~GHz, $\omega_{\mathbf{2}}/2\pi\approx 9.675$ GHz, and $\omega_{\mathbf{3}}/2\pi\approx 9.881$~GHz. We identify $f_{\mathbf{3}}$ as $R2$ and $f_{\mathbf{1}}$ and  $f_{\mathbf{2}}$ as spurious modes. (c) Electric field distribution from an eigenmode simulation for the spurious mode at $\omega_{\mathbf{1}}/2\pi$  assuming a 5~$\mu$m gap between the enclosure halves. From the distribution of the electric field we expect this mode to couple more strongly to $Q2$ than to $Q3$, around which the electric field is small. }
	\label{fig:3D_ramseySpect}
\end{figure}

The coherence spectroscopy method consists of repeated $T_2$ measurements performed while sweeping the frequency of a continuous wave (CW) tone that is applied to the qubit control port.   A change in $T_2$  is detected when the CW drive is resonant with a mode that couples to the qubit (see Fig. \ref{fig:Q1_echoSpect}). 
Here we apply coherence spectroscopy to two different devices. The first is a planar design with two qubits coupled by a bus resonator, where each qubit is measured through an individual coplanar waveguide resonator (Fig. \ref{fig:Q1_echoSpect}[a]). The second is a 3D microwave cavity device consisting of four qubits and five cavities (Fig. \ref{fig:3D_ramseySpect}[a]).  The $T_2$ measurement is accomplished by either a Ramsey or spin echo experiment depending on the device; although obtaining $T_2^*$ from Ramsey is generally sufficient for qubits with long coherence times, it can be preferable to use $T_2^\mathrm{echo}$ for qubit devices which still suffer from small amounts of residual charge noise\cite{schreier} ($\approx$~50-200~kHz).  For the results presented here, we measure $T_2^{\mathrm{echo}}$ for the planar device and $T_2^*$ for the 3D cavity experiments.

In the planar device, both qubit readout resonators have a frequency of 6.4 GHz and designed external quality factors in the range of 10,000-15,000.  The two qubits are coupled through a bus resonator, as shown in Fig. \ref{fig:Q1_echoSpect}(a). The qubit transition frequencies ($\omega_{01}/2\pi$) are 4.98 GHz and 5.17 GHz, and the measured  bus 
 resonator frequency is 7.24 GHz.  
 We apply control pulses to a single port, Port 1, to address the lower frequency qubit, $Q1$, and we use a spin echo experiment to measure $T_2$. 
Figure \ref{fig:Q1_echoSpect}(b) shows $T_2^{\mathrm{echo}}$ as a function of the CW drive frequency from 5 to 16 GHz.  The line at 6.4~GHz in this figure indicates the readout resonator, and  the second harmonic is visible at 12.8~GHz.  The bus resonator appears clearly at 7.24 GHz.  Weaker lines between 8 and 10 GHz indicate possible
spurious modes coupled to the qubit, but finer sweeps of the CW frequency over this range does not show any significant qubit dephasing.  In addition, there is a slight dip at the lower end  of this spectrum, which is due to the probe qubit coupling to the other qubit, $Q2$, on the sample with $\omega_{01}/2\pi =  5.17$ GHz.

Fitting to the $T_2^{\mathrm{echo}}$ versus CW frequency data provides several parameters that we otherwise cannot access.  
Ref. \onlinecite{gambetta_measurementInducedDephasing_2006} derives the theory behind the analytical description of $T_2^{\mathrm{echo}}$ used here as a function of the drive frequency and
strength in a qubit-cavity coupled system.  
The inverse of $T_2^{\mathrm{echo}}$ is the measurement induced dephasing rate, $\Gamma_m$, which is defined below,
\begin{eqnarray}
\Gamma_m &= \frac{D_s\kappa}{2}\\
\mbox{where }D_s &= \frac{2(\bar{n}_+ + \bar{n}_{-})\chi^2}{\kappa^2/4+\chi^2+\Delta^2_r}\\
\mbox{and }\bar{n}_{\pm} &= \frac{\epsilon_{\rm rf}^2}{\kappa^2/4 +(\Delta_r\pm\chi)^2} \label{eq:photonvspower}
\end{eqnarray}
Here $\Delta_r$ is the detuning of the applied drive from the bare cavity resonance, $\chi$ the dispersive shift of the cavity frequency
due to the state of the qubit, 
$n_{\pm}$ the average number of photons in the cavity when the qubit is in the ground (-) or excited (+) state, 
$\kappa$ the cavity decay rate, 
and $\epsilon_{\rm rf}$ the drive amplitude.   

We fit the data to the expression for $\Gamma_m$, allowing
$\kappa$, $\chi$, $\epsilon_{\rm rf}$, and the bare cavity frequency to be fit parameters. We also include a fit parameter to account for the effect of the qubit $T_1$ time, which causes an asymmetry in the $2\chi$ splitting of the resonance. As the qubit decays during the echo experiment, the larger amplitude peak indicates the pull of the cavity when the qubit is in the ground state. Overall, the fit to theory in Fig \ref{fig:Q1_echoSpect}(c) is in good agreement with the data.

From the fitted parameters we are able to calculate several additional quantities, including the quality factor of the bus resonator (found from $\kappa$) and the 
average number of photons in the cavity (using Eq. \ref{eq:photonvspower}).  The quality factor for the bus is measured to be on the order of 100,000.  This high-Q illuminates the difficulty of identifying these modes through S parameter measurements. We have plotted in the inset of Fig. \ref{fig:Q1_echoSpect} (b) the photon number versus the power applied at the input port for both resonators coupled to the qubit.  The number of photons in the cavity is linear with applied power, showing agreement with Eq. \ref{eq:photonvspower}.


In a well-controlled coplanar waveguide system where proper care has been taken to connect the different regions of the ground plane we find no evidence for the existence of spurious modes. This is typically also true for superconducting qubits enclosed in a single 3D waveguide cavity. The waveguide cavity provides a very clean microwave environment and allows for qubit lifetimes exceeding 100 $\mu s$ \cite{paik,Rigetti2012,dial}. In addition to long life- and  coherence times, the 3D approach provides flexibility in optimizing qubit properties by pre-screening the qubits before qubit/cavity assembly. The difficulty lies in building multi-qubit 3D systems. One solution is to stretch the qubit between two cavities, as shown in Fig. \ref{fig:3D_ramseySpect}(a), so that multiple qubits are coupled through a central coupling cavity and addressed through independent readout cavities. We refer to such qubits as ``bridge qubits''\cite{kirchmair}. 

We typically find a significant reduction in coherence times of bridge qubits compared to qubits placed in single cavities. This reduced coherence may be the result of a larger number of spurious modes supported by the multi-cavity enclosures. The larger mating surface area increases the demands on the machining tolerances in order to have a close fit over the entire piece. Additionally, each cavity is designed with thin walls ($\approx$ 0.4 mm wide) to allow the qubit to extend into neighboring cavities while maintaining strong  coupling to both cavities. Misalignment and machining imperfections can easily lead to vacuum gaps between separate pieces of the enclosure, which can in turn support spurious microwave modes. Extra channels of relaxation and decoherence are expected if the spurious modes couple to the qubit mode. 

We extend the method of coherence spectroscopy to characterize spurious modes in the 3D multi-cavity and multi-bridge qubit system. The 3D device consists of four bridge qubits and five waveguide cavities (Fig. \ref{fig:3D_ramseySpect}[a]). The four qubits are coupled to a common bus cavity and are measured through individual readout cavities, all of which are machined from a single aluminum block. The frequencies of the qubits ($Q$1-$Q4$) are $\omega_{Q1}/2\pi= 4.677$ GHz, $\omega_{Q2}/2\pi= 4.716$ GHz, $\omega_{Q3}/2\pi= 4.851$ GHz, and $\omega_{Q4}/2\pi= 4.865$ GHz. The dressed readout cavity frequencies are $\omega_{R1}/2\pi= 10.131$ GHz, $\omega_{R2}/2\pi= 9.882$ GHz, $\omega_{R3}/2\pi= 10.128$ GHz, and $\omega_{R4}/2\pi= 10.858$ GHz and the bus resonance is $\omega_{B}/2\pi=6.9698$ GHz. As in the case of the tested planar device, direct scattering parameter measurements do not give any indications of spurious modes, in part due to the strong filtering of the transmitted signal by the cavities, and in part due to heavy attenuation of the drive lines that is required to minimize the number of thermal photons that reach the cavity. 

Using $Q2$ as our coherence spectroscopy sensor, we apply a CW tone to the main bus cavity and search for spurious modes in our system through observations of Ramsey experiments on $Q2$. We observe regions in the Ramsey spectrum with sharp shifts in frequency, as seen in Fig. \ref{fig:3D_ramseySpect} (b). In a scan over the frequency range 7.5 GHz to 10.2 GHz we observe three such features located at $\omega_{1}/2\pi\approx 7.805$ GHz, $\omega_{2}/2\pi\approx 9.675$ GHz, and $\omega_{3}/2\pi\approx 9.881$ GHz. We readily identify the feature at $\omega_{3}/2\pi$ as the readout resonator $R2$. The features at $\omega_{1}/2\pi$ and $\omega_{2}/2\pi$ do not, however, correspond to any designed modes in the system and are determined to be spurious modes. 

In order to gain a better understanding of the origin of these modes, we use a microwave eigenmode solver (Ansoft HFSS) and model the gap between the two halves of the enclosure as a 5~$\mu$m separation.  The simulation shows that modes can be confined in this separation along the surface of the cavity walls. Using this model, we predict a mode at 7.838 GHz (See Fig.\ref{fig:3D_ramseySpect}(c)), with an electric field distribution which overlaps $Q2$. This mode is in good agreement with the mode found at $\omega_1/2\pi$ via coherence spectroscopy. The simulation also indicates a mode at 9.873 GHz, which we identify as the feature at $ \omega_2/2\pi$.  The ability to distinguish this second spurious mode is complicated by the presence of the readout resonator at 9.882~GHz, which is closer to the simulated frequency than $ \omega_2/2\pi$, but $ \omega_2/2\pi$ is approximately 2\% error from $ \omega_2/2\pi$  of the simulated value and we do not expect higher accuracy from the microwave simulations.  

 The strong agreement in frequency between simulations and measurements suggests that the measured modes are largely related to the vacuum gap between the enclosure halves.  We also note that we find additional modes in simulation (7.535 GHz for instance) that we do not observe in coherence spectroscopy, likely because they do not have significant electric field components around $Q2$ and thus have low coupling to the qubit (see Fig.~\ref{fig:3D_ramseySpect}[c]).

Both in 2D and 3D superconducting qubit architectures, we have demonstrated coherence spectroscopy, a tool that provides the ability to detect both high-Q microwave resonators that couple weakly to device ports as well as spurious modes arising from package geometry and interfaces. This characterization technique provides estimates of the coupling and Q of microwave modes, allowing for design parameter verification.  Additionally, by comparison with microwave simulation we are able to determine the source of spurious modes that cause dephasing, which will allow for improvements to the qubit environment that should lead to longer coherence times.  Unlike S-matrix measurements, coherence spectroscopy detects only those modes that couple to the qubit, thus discriminating the most relevant modes for device performance from all that may appear on a network analyzer.  

Superconducting qubit systems are becoming more complex, with larger numbers of qubits and resonators and varying device structures\cite{kelly15, qx3,versluis}.  Indeed, many proposed paths forward involved hybrid systems of 2D qubits and 3D structures \cite{minev}.  This complexity will necessitate careful design for microwave hygiene, and characterization methods like coherence spectroscopy will be critical in these efforts. 

We acknowledge George A. Keefe, Mary B. Rothwell and Jim Rozen for experimental contributions and David C. McKay for comments on the manuscript. This work was supported by ARO under contract W911NF-14-1-0124 and IARPA under contract W911NF-10-1-0324.

\bibliography{EchoSpectPaper}

\end{document}